# Fault-Tolerant Design Approach Based on Approximate Computing

P. Balasubramanian, D.L. Maskell
School of Computer Science and Engineering
Nanyang Technological University
Singapore 639798
Correspondence: balasubramanian@ntu.edu.sg

## Abstract
Triple Modular Redundancy (TMR) has been traditionally used to ensure complete tolerance to a single fault or a faulty processing unit, where the processing unit may be a circuit or a system. However, TMR incurs more than 200% overhead in terms of area and power compared to a single processing unit. Hence, alternative redundancy approaches were proposed in the literature to mitigate the design overheads associated with TMR, but they provide only partial or moderate fault tolerance. This research presents a new fault-tolerant design approach based on approximate computing called FAC that has the same fault tolerance as TMR and achieves significant reductions in the design metrics for physical implementation. FAC is suited for a plethora of error-tolerant applications. Here, the performance of TMR and FAC has been evaluated for a digital image processing application. The image processing results obtained confirm the usefulness of FAC. When an example processing unit was implemented using a 28-nm CMOS technology, FAC achieved a 15.3% reduction in delay, a 19.5% reduction in area, and a 24.7% reduction in power compared to TMR.

## 1. Introduction
Due to the reduction in the size of transistors, processing units comprising them like electronic circuits and systems become more susceptible to faults or failures during regular operation [1][2] or aging [3]. This susceptibility increases when processing units are subjected to challenging conditions, such as space, where the occurrence of high-energy radiation [4] is given. Radiation can have various sources and causes when it comes to harsh environments, and various sources of radiation can lead to different types of radiation effects that could impact the performance, reliability, and functionality of electronic components. The types of radiation effects on electronic devices include single-event effects, total ionizing dose effects, displacement damage, and more. These effects can cause temporary or permanent changes in electronic behavior such as single-event upsets, latch-ups, gate ruptures, and increased leakage currents. Therefore, the design of electronic circuits and systems to withstand these radiation effects involves using radiation-hardened materials, shielding, redundancy, error correction codes, etc. This work focuses on redundancy as a fault tolerance design strategy to address faults that may arise in processing units within a specified limit due to the impact of radiation. In this context, a new fault-tolerant design approach involving approximate computing has been proposed. The proposed approach is suitable for the redundant implementation of data path logic which is error-tolerant. Indeed, many practical applications are inherently error-resilient and these include multimedia encompassing digital signal processing, big data analytics, neuromorphic computing, hardware realization of deep neural networks for machine learning and artificial intelligence, memory systems for multicore processors, computer graphics and vision, and ultra-low power electronic design involving the sub-threshold operation of devices, etc. The potential of approximate computing to achieve greater energy efficiency at the cost of an acceptable compromise in accuracy has already been demonstrated in the literature.

## 2. Survey of Related Work
N-Modular Redundancy (NMR) is a well-known approach that uses N identical processing units, where N is odd (i.e., N = 3 or more). The corresponding outputs of the N processing units are combined using majority voters to generate the final output. In NMR, given N identical processing units, the majority, i.e., (N + 1)/2 processing units must function correctly to ensure the proper functioning of the NMR scheme, assuming the majority voter also operates correctly. Nevertheless, the majority voter may be hardened like a processing unit by duplicating it to ensure the correct operation. According to NMR, faults of a maximum of (N − 1)/2 processing units can be tolerated without affecting the final output.

Triple Modular Redundancy (TMR) represents the fundamental version of NMR and enjoys widespread popularity and use. In a practical study [7], various Virtex FPGA devices were exposed to radiation from protons and heavy ions. The study revealed that single-bit upsets accounted for 96% to 99% of all upsets, with multiple-bit upsets making up the remaining percentage. Considering the dominant prevalence of single-bit upsets in high-energy radiation environments like space, TMR offers an effective solution. TMR involves the use of three identical processing units, and their respective outputs are combined through majority voting to produce the primary output. Consequently, TMR can tolerate any single fault or any faulty processing unit. However, implementing TMR requires two additional (identical) processing units and a majority voting logic compared to a single processing unit. As a result, a TMR implementation incurs about 200% overheads in terms of area and power compared to a single processing unit.

To mitigate the area and power overheads associated with TMR, researchers have proposed compromise approaches [8][9][10][11] that aim to minimize design metrics such as area, power, and delay while compromising on fault tolerance. One such approach is the Selective insertion of TMR (STMR) [8]. STMR suggests applying TMR only to the critical components of a processing unit while leaving the less critical parts as a simple implementation. By adopting STMR instead of the conventional full TMR, it becomes possible to reduce both the area and power requirements of a redundant implementation. However, there are a couple of challenges associated with STMR. Firstly, determining which parts of a processing unit are critical and which are not may not be straightforward for all practical applications. Moreover, this differentiation may not remain valid throughout the operational lifetime of a processing unit. Secondly, if the unprotected, less critical parts of a processing unit are affected, there is no guarantee that the outputs of the processing unit will remain unaffected.

Subsequently, the concept of Approximate TMR (ATMR) [9] was introduced. ATMR involves using one accurate processing unit and two different approximate processing units with reduced logic. The outputs of the accurate and approximate processing units are majority-voted to generate the primary outputs. Unlike traditional TMR which utilizes three accurate processing units, ATMR enables reduced area and power dissipation since it has one accurate and two approximate processing units. However, the implementation of ATMR comes with certain challenges. Firstly, if either the accurate processing unit or one of the approximate processing units produces a faulty output, the corresponding output of ATMR could become erroneous. Secondly, if the accurate processing unit itself becomes faulty, and its outputs do not match the outputs of the approximate units, ATMR could experience failure. These scenarios indicate that ATMR is not fully tolerant to a single fault or a faulty processing unit, which goes against the fundamental property of TMR (i.e., the ability to reliably mask any single fault or a faulty processing unit).

An alternative approach called Fully Approximate TMR (FATMR) was also introduced [9][10] to address the design overheads associated with traditional TMR. FATMR employs three distinct approximate versions of an original accurate processing unit whose corresponding outputs are subjected to majority voting using accurate majority voters. In FATMR, the outputs of any two approximate processing units align, meaning that if one of the approximate units produces a faulty output, the corresponding output of FATMR would be erroneous. Moreover, if any of the approximate processing units were to become faulty, it could jeopardize the FATMR implementation, leading to inaccurate outputs. Consequently, FATMR tends to exhibit a higher degree of unreliability compared to ATMR. Both ATMR and FATMR are unsuitable for safety-critical applications due to the inherent uncertainty in their output, even in the presence of a single fault or a faulty processing unit. Further, to the best of our knowledge, there is no demonstration of the usefulness of ATMR and FATMR in any practical application.

A novel redundancy technique called Majority Voting-based Reduced Precision Redundancy (MVRPR) was introduced [11], specifically targeting naturally error-resilient applications. One such application is digital signal processing, which encompasses tasks like digital image, video, and audio processing, etc. These applications inherently possess a degree of error tolerance since minor distortions in images or videos or subtle background noise in audio might not be discernible to the human eye or ear due to the limitations of human perception. By reducing the precision of the digital system, it becomes possible to lower its design metrics and enhance its energy efficiency, making MVRPR relevant in this context. The authors [11] focused on describing the design of an MVRPR adder. The key feature of an MVRPR adder is its division into two equal-sized parts: a significant part and a less significant part. The categorization of these parts as significant and less significant is based on the weightage assigned to the sum bits generated by each respective part. The sum bits from both parts are however concatenated to obtain the final sum output. In the MVRPR implementation, the significant part of the adder benefits from TMR protection. This means that the significant part is triplicated, and its corresponding outputs are subjected to majority voting, ensuring high reliability. On the other hand, the less significant part remains as a single, unprotected implementation. As a result, any potential single-bit upset(s) that affect the less significant part could affect the primary output. Because of the absence of protection for the less significant part of MVRPR, its fault tolerance capability is only moderate compared to TMR [11][12].

## 3. Proposed Redundancy Approach – FAC

We presented a novel **F**ault-tolerant design approach based on **A**pproximate **C**omputing, abbreviated as **FAC** [6]. Before delving into the details of FAC, we briefly discuss the merits of approximate computing. Approximate computing is a promising alternative to traditional accurate computing, especially for inherently error-tolerant applications. By accepting a certain level of compromise on the computation accuracy, approximate computing offers advantages such as reduced area, lower power dissipation, higher processing speed, and improved energy efficiency [13][14]. The benefits of approximate computing have been demonstrated for various practical applications, particularly those that exhibit inherent error resilience, such as multimedia encompassing digital signal processing, computer graphics, computer vision, neuromorphic computing, and the implementation of hardware for AI, machine learning, and neural networks, etc. [15]. Consequently, leveraging the potential of approximate computing becomes an appealing prospect for designing fault-tolerant processing units, especially in resource-constrained environments like space, where area efficiency, low power dissipation, high processing speed, and energy efficiency are critical factors.

Previous works [8][9][10][11] have suggested alternative (approximate/accurate) implementations of redundancy. However, as discussed in Section 2, these alternative approaches suffer from drawbacks and are unlikely to be used for practical applications. In contrast, the proposed FAC could potentially be a practical alternative. FAC is generic and can be applied to address any order of NMR. Nonetheless, here, we consider a 3-tuple version of FAC to facilitate a direct comparison with TMR. To showcase the distinction between TMR and FAC, we illustrate their representative architectures in Figures 1a and b respectively.

In TMR, three identical processing units are used, as shown in Figure 1a, and the processing units are all accurate. The majority voters utilized in TMR are also accurate. The outputs of each processing unit are represented by A and B, with output A assumed to have more significance than output B. This assumption is reasonable, particularly for arithmetic circuits, where the output bits vary in significance from most significant to least significant. The outputs A and B from the processing units are subjected to voting using (accurate) majority voters 1 and 2, respectively. A 3-input majority voter synthesizes the Boolean function F = XY + YZ + XZ, where F denotes the output, and X, Y, and Z denote the inputs. Various majority voter designs relevant to TMR are available [16][17], with the majority voter typically assumed to be perfect. However, if the majority voter cannot be assumed to be perfect, redundancy can be applied to the majority voter, like the processing units. The primary outputs of the TMR implementation, denoted as V1 and V2, are the outputs of majority voters 1 and 2, respectively. By triplicating the processing units and using majority voters, a TMR implementation would effectively conceal a single fault(s) or a faulty processing unit.

Referring to Figure 1b, the proposed redundancy approach (FAC) involves partitioning a processing unit into two parts based on the significance of their corresponding outputs in relation to the primary output. FAC involves dividing the processing unit into two tailored parts based on an application's requirement. This implies that the two parts could be of equal or unequal size, depending upon the application. FAC triplicates both the significant and less significant parts. Further, in FAC, the less significant part of the processing unit is approximated instead of being retained accurately. The constituents of processing units 1, 2, and 3 are depicted within red, blue, and brown boxes in dashed lines in Figure 1b. Nevertheless, all three processing units are identical. It should be noted that each processing unit's less significant part in FAC is identical and may or may not be connected to its corresponding significant part. The connection depends on the manner of logical approximation applied to the less significant part of the processing unit. The connections between the less significant and significant parts of each processing unit in FAC are represented by dotted black lines in Figure 1b, with the intermediate output being denoted as T. FAC features the same fault tolerance as TMR, as it can mask any single fault in the significant or less significant part of any processing unit or tolerate a faulty processing unit. However, because the triplicated less significant parts are approximated, the practical applicability of FAC depends on two factors: (i) The manner of logical approximation applied to the less significant parts of the processing unit, (ii) The extent of approximation incorporated in these less significant parts.

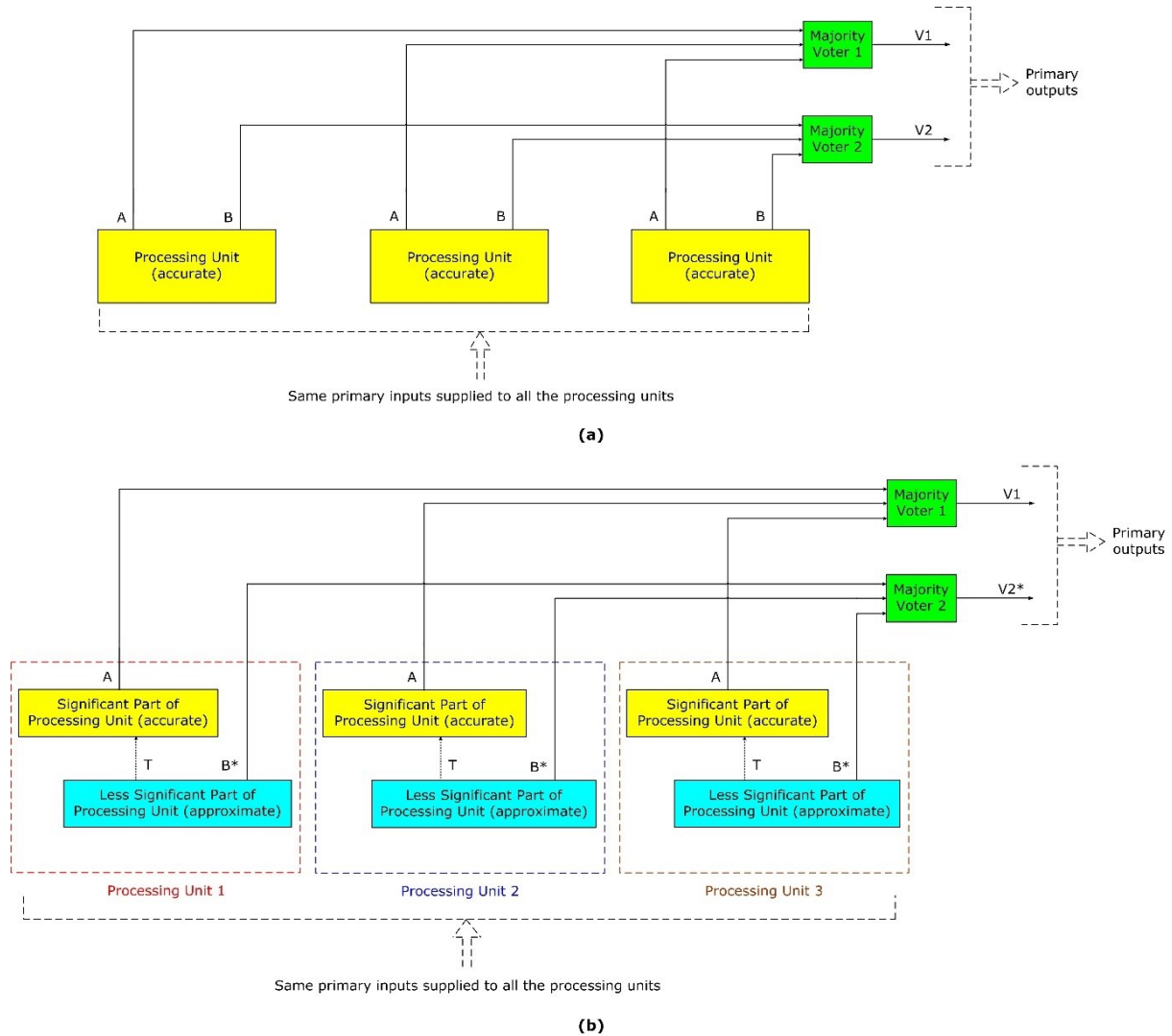

**Figure 1.** Example block-level illustrations of (a) TMR architecture, and (b) proposed FAC architecture. B* of (b) may or may not be equal to B of (a), depending upon the inputs supplied.

In Figure 1b, output A from processing units 1, 2, and 3 is subjected to voting using majority voter 1, resulting in output V1. This voting mechanism is also employed in TMR. As mentioned earlier, the triplicated less significant parts of processing units 1, 2, and 3 in FAC are identical but approximate. Consequently, the output B* from processing units 1, 2, and 3 may or may not be equivalent to the output B of the accurate processing unit, and this depends on the inputs supplied. For instance, if we assume that a 2-input EXOR gate having inputs X and Y was used to generate output B in TMR, the EXOR gate will output 1 if X ≠ Y and 0 if X = Y. If, due to approximation, the 2-input EXOR function (of TMR) is replaced by a 2-input OR function in FAC, the OR gate will output 1 when X = Y = 1, and X ≠ Y, and output 0 only when X = Y = 0. Thus, for the conditions where X = Y = 0 and X ≠ Y, both EXOR and OR gates will produce the same output; however, when X = Y = 1, the outputs of the two gates will differ. Hence, B* may or may not be equal to B based on the inputs supplied. In FAC, the output B* from the less significant parts of processing units 1, 2, and 3 are subjected to voting using majority voter 2, resulting in the output V2*. It should be noted that V2* may or may not be equal to V2, and V1 and V2* represent the primary outputs of the FAC implementation.

Arithmetic circuits, including adders, multipliers, dividers, and data paths containing functions like the discrete Cosine transform, finite/infinite impulse response filter, and sum of absolute difference, etc. exhibit varying degrees of significance in their output bits. This characteristic allows the partitioning of such processing units into significant and less significant parts, thus presenting an opportunity for implementing them according to FAC. Depending on the target application, the less significant part of a processing unit can be approximated to a suitable degree. Similarly, logic functions can be redundantly implemented according to FAC, and again, the level of logic approximation for the less significant part should be determined based on the target application.

## 4. Application – Digital Image Processing

To compare the performance of TMR and the proposed FAC, a digital image processing case study involving fast Fourier transform (FFT) and inverse fast Fourier transform (IFFT) [18] was considered. A set of 8-bit grayscale images with a spatial resolution of 512 × 512 was randomly chosen for evaluation. Each image was converted into a matrix format and subjected to FFT computation, followed by image reconstruction using IFFT. The FFT and IFFT computations were carried out in integer precision with scaling[19] to ensure that no data loss or overflow occurred during the computations. Multiplication was performed accurately, while the addition was performed accurately using the precise adder and inaccurately using an imprecise adder, separately. The architecture of the imprecise adder [20] used in the FAC approach is depicted in Figure 2 which contains two parts viz. a violet section representing the accurate part and a pink section representing the approximate part. The accurate and approximate parts of the imprecise adder are marked in Figure 2 for easy comparison with Figure 1b. The accurate part is considered significant, while the approximate part is regarded as less significant. In Figure 2, the adder size is N bits, the size of the approximate adder part is L bits, and the size of the accurate adder part is (N–L) bits. The adder inputs are represented by $A_{N-1}$ up to $A_0$ and $B_{N-1}$ up to $B_0$, while the adder output is denoted by $SUM_N$ up to $SUM_0$. Subscripts (N–1) and 0 indicate the most significant and least significant bit of the adder inputs, and subscripts N and 0 signify the most significant and least significant bit of the adder's sum outputs, respectively. The accurate part adds (N–L) input bits along with a carry input provided by the approximate part and produces (N–L+1) sum bits. In the approximate part, sum bits $SUM_{L-1}$ up to $SUM_{L-4}$ have reduced logic while the remainder of the sum bits $SUM_{L-5}$ up to $SUM_0$ are assigned a constant 1 (binary). For synthesis, $SUM_{L-5}$ up to $SUM_0$ are individually connected to tie-to-high standard library cells. The value of L is typically determined based on the maximum error tolerable for a given application.

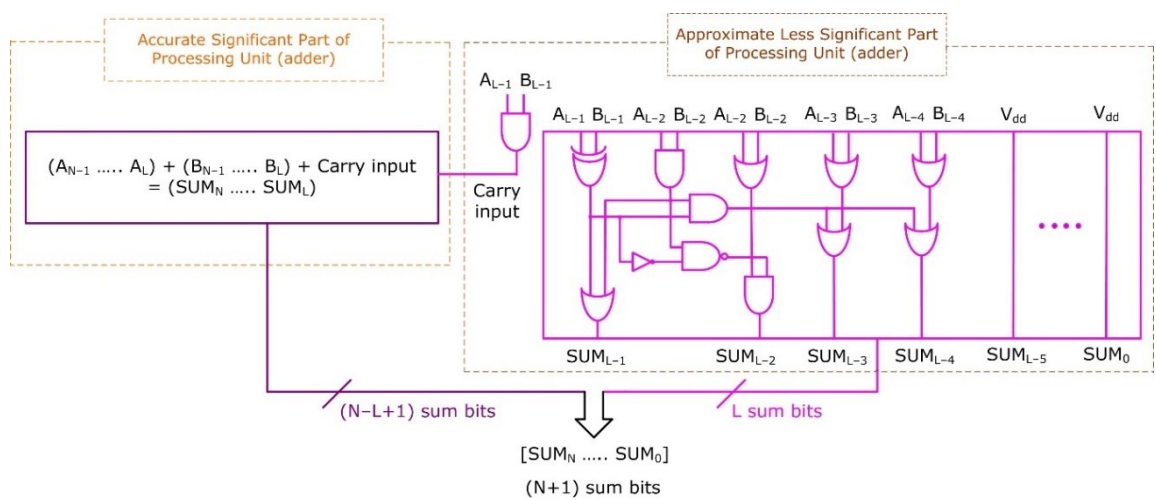

**Figure 2.** The architecture of the N-bit imprecise adder used for FAC implementation in this work.

TMR utilizes the precise adder while FAC uses an imprecise adder (illustrated in Figure 2). Experimentation with many images was conducted to determine the maximum allowable approximation for the imprecise adder, ensuring acceptable image quality after processing. The fundamental principle in approximate computing is incorporating the highest degree of approximation that maintains an acceptable level of output quality. Typically, the nature and/or extent of approximation are determined through trial and error specific to a given application. From a hardware standpoint, employing less approximation than an application can accommodate called 'under-approximation' would yield satisfactory output quality but curtail the potential savings in design metrics achievable when compared to using precise hardware. Conversely, adopting a level of approximation that exceeds what an application can tolerate called 'over-approximation' would result in subpar and unsatisfactory output quality (despite yielding exaggerated savings in design metrics), which is not desirable. Hence, the ideal approach is to identify the 'optimum approximation' that an application can embed while ensuring a practically acceptable output quality. This strategy allows for the maximization of design metric savings compared to using precise hardware, without compromising the output quality beyond the acceptable threshold for the given application. In a prior work[21], it has been illustrated how the quality of processed digital images varies for the three example scenarios of under-approximation, optimum approximation, and over-approximation. Here, for a 32-bit addition, the maximum acceptable approximation while ensuring an acceptable output quality (here, image quality) was found to be the use of 10 sum bits for the approximate adder part (L = 10) and 22 sum bits for the accurate adder part (N–L = 22), which was determined based on trial and error.

The results of digital image processing corresponding to TMR and FAC are depicted in Figure 3. To assess the quality of the processed images, the Peak Signal-to-Noise Ratio (PSNR)[22] and the Structural Similarity Index Measure (SSIM)[23] were calculated. PSNR serves as a general figure of merit for digital signal processing while SSIM specifically measures digital image processing quality. The ideal values are PSNR = ∞ and SSIM = 1 (decimal). In Figure 3, the images processed using the accurate adder representative of TMR exhibit ideal values of PSNR and SSIM. Despite using an approximate adder for the less significant part, the FAC implementation is found to consistently produce high-quality images comparable to the accurate (TMR) implementation.

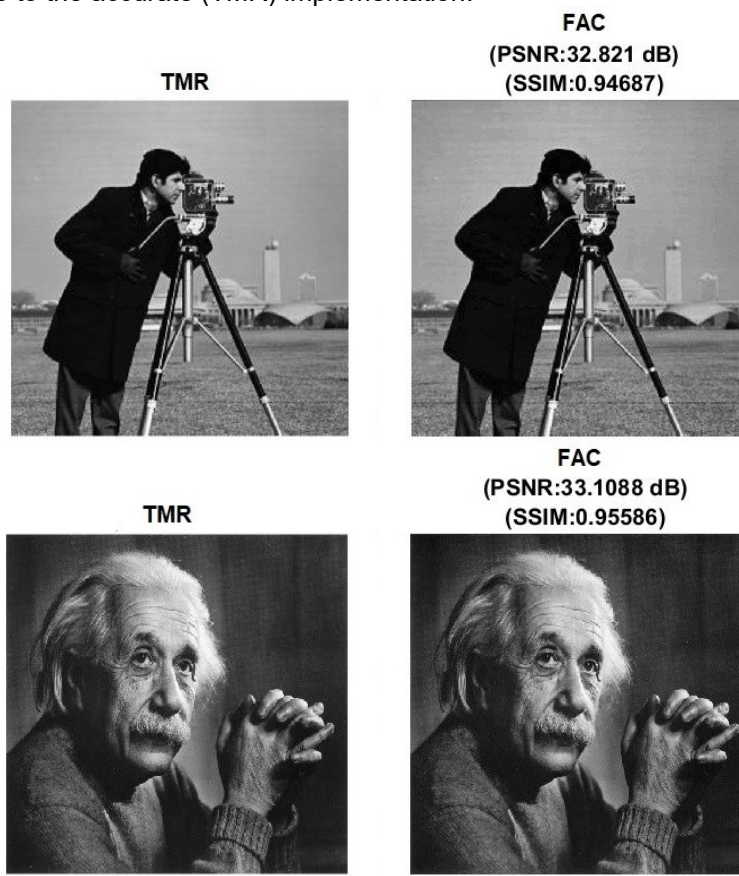

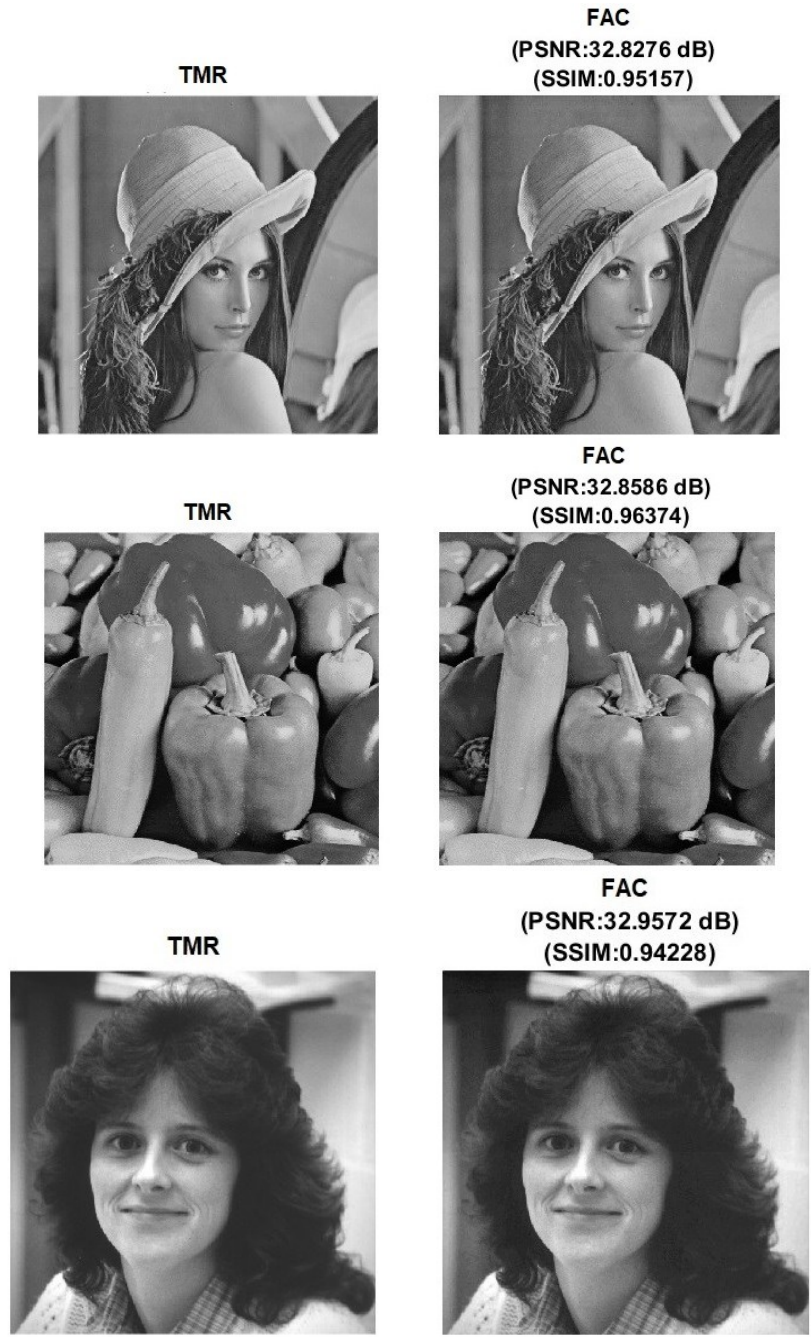

**Figure 3.** Results of digital image processing corresponding to TMR and the proposed FAC based on experimentation with some images. PSNR = ∞ and SSIM = 1 for the images processed according to TMR, which are ideal. FAC consistently yield images with PSNR > 30 dB and SSIM close to 1, which are acceptable.

## 5. Implementation and Design Metrics

To physically implement the adders used for digital image processing (since the adder alone differs between TMR and FAC implementations), we structurally described each of them in Verilog HDL. The adders considered are (i) An accurate 32-bit carry-lookahead adder (CLA) [24] for the simplex implementation, (ii) A 32-bit TMR adder, and (iii) A 32-bit FAC adder having a 22-bit significant part and a 10-bit less significant part (since the 22-10 input partition was found to be optimum for digital image processing, as noted in the previous section). The TMR adder utilized the accurate CLA structure [24], and the accurate part of the FAC adder was also realized based on the same CLA structure. All adders were synthesized using a 28-nm CMOS standard digital cell library [25]. A typical low-leakage library specification featuring a 1.05 V supply voltage and a 25 °C operating junction temperature was considered. During simulation and synthesis, default wire load and a fanout-of-4 drive strength were assigned to all sum bits. Synopsys EDA tools were used for synthesis, simulation, and the estimation of design metrics. Design Compiler was used for synthesis and to estimate the total area of the adders, including cells and interconnect area. To evaluate the performance of the adders, a test bench comprising over one thousand random inputs was supplied at a latency of 2 ns (500 MHz) to simulate their functionality using VCS. The switching activity was recorded while performing functional simulation, which was used to estimate the total power dissipation using Prime Power. Prime Time was used to estimate the critical path delay of each implementation. The design metrics of the adders, including area, power dissipation, and critical path delay, are given in Table 1.

**Table 1.** Design metrics of non-redundant and redundant adders, implemented using a 28-nm CMOS process technology.

| Implementation | Area (µm$^2$) | Delay (ns) | Power (µW) |
|---|---|---|---|
| Single adder (CLA) | 527.45 | 1.13 | 91.7 |
| TMR adder | 1752.43 | 1.24 | 291.8 |
| FAC adder | 1410.06 | 1.05 | 219.8 |

The single adder (i.e., accurate CLA) exhibits the lowest area, and power dissipation among all the adders considered, but is not fault-tolerant. In comparison, the TMR adder experiences increased delay due to the additional majority voter delay in its critical data path. The TMR adder occupies 2.3× more area and dissipates 2.2× more power compared to the single adder since it includes two extra CLAs and a majority voting logic. The proposed FAC adder features a 22-bit accurate (significant) part, and its critical path is governed solely by this accurate part. This is because, as shown in Figure 2, the accurate and approximate FAC adder parts are connected by a small carry input logic (represented by the internal output T in Figure 1b), defined as the logical conjunction of input bits $A_{L-1}$ and $B_{L-1}$. As a result, the FAC adder exhibits a reduced critical path delay even compared to the single adder (accurate CLA), and the TMR adder. Additionally, the approximate 10-bit sum logic of the FAC adder results in a smaller silicon footprint, leading to reduced power dissipation in comparison to the TMR adder. The proposed FAC adder reports significant reductions in design metrics compared to the TMR adder, including a 15.3% reduction in delay, a 19.5% decrease in area, and a 24.7% reduction in power.

Two commonly used metrics for assessing a digital logic design's energy efficiency and design efficiency are the power-delay product (PDP) and the power-delay-area product (PDAP). Minimizing power, delay, and area is desirable, so it follows that minimizing PDP and PDAP is also desirable. Compared to the TMR adder, the FAC adder achieves a 36.2% reduction in PDP and a 48.7% reduction in PDAP. From Figure 3 and Table 1, it is inferred that FAC is preferable to TMR for implementing the data path of inherently error-tolerant applications.

## 6. Conclusions

A novel fault-tolerant design approach called FAC was introduced which has the same fault tolerance as NMR. Specifically, a 3-tuple version of FAC was examined, allowing for a direct comparison with TMR. A digital image processing application (representative of an error-tolerant application) was considered as the case study and the results obtained demonstrate the usefulness of FAC. For the example implementation considered, FAC was found to achieve reductions in all design metrics compared to TMR, without compromising the fault tolerance. While FAC may be useful for data paths, it may not be suitable for the implementation of control logic due to the general requirement of 100% correctness which is better realized using accurate logic. To cope with multiple faults (i.e., with more than one corresponding output bit affected or more than one processing unit failing), according to NMR, a higher-order version such as quintuple modular redundancy, septuple modular redundancy, etc., may have to be used; the corresponding equivalent according to the proposed architecture would be a 5-tuple version of FAC, a 7-tuple version of FAC, etc.

**Acknowledgment**
This research was partially funded by the Singapore Ministry of Education (MOE), Academic Research Fund under grant numbers Tier-1 RG48/21 and Tier-1 RG127/22.

**References**

1. Miskov-Zivanov, N.; Marculescu, D. Multiple transient faults in combinational and sequential circuits: A systematic approach. IEEE Transactions on Computer-Aided Design of Integrated Circuits and Systems, 2010, 29, 1614–1627.

2. Baumann, R.C. Radiation-induced soft errors in advanced semiconductor technologies. IEEE Transactions on Device and Materials Reliability, 2005, 5, 305–316, .

3. Rossi, D.; Omana, M.; Metra, C.; Paccagnella, A. Impact of aging phenomena on soft error susceptibility. In Proceedings of the IEEE International Symposium on Defect and Fault Tolerance in VLSI and Nanotechnology Systems, Vancouver, BC, Canada, 3–5 October 2011.

4. Mahatme, N.N.; Bhuva, B.; Gaspard, N.; Assis, T.; Xu, Y.; Marcoux, P.; Vilchis, M.; Narasimham, B.; Shih, A.; Wen, S.-J. Terrestrial SER characterization for nanoscale technologies: a comparative study. In Proceedings of the IEEE International Reliability Physics Symposium, Monterey, CA, USA, 19–23 April 2015.

5. Balasubramanian, P.; Maskell, D.L. A fault-tolerant design strategy utilizing approximate computing. In Proceedings of the IEEE Region 10 Symposium (TENSYMP), Canberra, Australia, 6–8 September 2023.

6. Balasubramanian, P.; Maskell, D.L. FAC: A fault-tolerant design approach based on approximate computing. Electronics, 2023, 12, Article #3819.

7. Quinn, H.; Graham, P.; Krone, J.; Caffrey, M.; Rezgui, S. Radiation-induced multi-bit upsets in SRAM-based FPGAs. IEEE Transactions on Nuclear Science, 2005, 52, 2455–2461.

8. Gomes, I.A.C.; Martins, M.G.A.; Reis, A.I.; Kastensmidt, F.L. Exploring the use of approximate TMR to mask transient faults in logic with low area overhead. Microelectronics Reliability, 2015, 55, 2072–2076.

9. Arifeen, T.; Hassan, A.S.; Moradian, H.; Lee, J.A. Input vulnerability-aware approximate triple modular redundancy: Higher fault coverage, improved search space, and reduced area overhead. Electronics Letters, 2019, 54, 934–936.

10. Ruano, O.; Maestro, J.A.; Reviriego, P. A methodology for automatic insertion of selective TMR in digital circuits affected by SEUs. IEEE Transactions on Nuclear Science, 2009, 56, 2091–2102.

11. Ullah, A.; Reviriego, P.; Pontarelli, S.; Maestro, J.A. Majority voting-based reduced precision redundancy adders. IEEE Transactions on Device and Materials Reliability, 2018, 18, 122–124.

12. Balasubramanian, P. Analysis of redundancy techniques for electronics design—Case study of digital image processing. Technologies, 2023, 11, Article #80.

13. Han, J.; Orshansky, M. Approximate computing: An emerging paradigm for energy-efficient design. In Proceedings of the 18th IEEE European Test Symposium, Avignon, France, 27–31 May 2013.

14. Venkataramani, S.; Chakradhar, S.T.; Roy, K.; Raghunathan, A. Approximate computing and the quest for computing efficiency. In Proceedings of the 52nd ACM/EDAC/IEEE Design Automation Conference, San Francisco, CA, USA, 8–12 June 2015.

15. Mittal, S. A survey of techniques for approximate computing. ACM Computing Surveys, 2016, 48, 1–33.

16. Balasubramanian, P.; Mastorakis, N.E. Power, delay and area comparisons of majority voters relevant to TMR architectures. In Proceedings of the 10th International Conference on Circuits, Systems, Signal and Telecommunications, Barcelona, Spain, 13–15 February 2016.

17. Balasubramanian, P.; Prasad, K. A fault tolerance improved majority voter for TMR system architectures. WSEAS Transactions on Circuits and Systems 2016, 15, 108–122.

18. Zhu, N.; Goh, W.L.; Zhang, W.; Yeo, K.S.; Kong, Z.H. Design of low-power high-speed truncation-error-tolerant adder and its application in digital signal processing. IEEE Transactions on VLSI Systems, 2010, 18, 1225–1229.

19. Nayar, R.; Balasubramanian, P.; Maskell, D.L. Hardware optimized approximate adder with normal error distribution. In Proceedings of the IEEE Computer Society Annual Symposium on VLSI, Limassol, Cyprus, 6–8 July 2020.

20. Balasubramanian, P.; Nayar, R.; Maskell, D.L. An approximate adder with reduced error and optimized design metrics. In Proceedings of the IEEE Asia Pacific Conference on Circuits and Systems, Penang, Malaysia, 22–26 November 2021.

21. Balasubramanian, P.; Nayar, R.; Maskell, D.L.; Mastorakis, N.E. An approximate adder with a near-normal error distribution: Design, error analysis and practical application. IEEE Access 2021, 9, 4518–4530.

22. Bovik, A. Handbook of Image and Video Processing, 2nd edition; Academic Press: Cambridge, MA, USA, 2005.

23. Zhou, W.; Bovik, A.C.; Sheikh, H.R.; Simoncelli, E.P. Image quality assessment: From error visibility to structural similarity. IEEE Transactions on Image Processing, 2004, 13, 600–612.

24. Balasubramanian, P.; Mastorakis, N.E. High-speed and energy-efficient carry look-ahead adder. Journal of Low Power Electronics and Applications, 2022, 12, Article #46.

25. Synopsys SAED_EDK32/28_CORE Databook. Revision 1.0.0. January 2012.